\documentclass[aps,preprint,preprintnumbers,amsmath,amssymb,amsfonts,groupedaddress,superscriptaddress,showpacs,showkeys,floatfix]{revtex4-1}

\usepackage{amsmath}
\usepackage{amssymb}
\usepackage{bm}
\usepackage{graphicx}
\usepackage{xcolor}
\usepackage{epstopdf}

\newcommand{\bmath}{\begin{mathletters}}
\newcommand{\emath}{\end{mathletters}}
\newcommand{\be}{\begin{eqnarray}}
\newcommand{\ee}{\end{eqnarray}}
\newcommand{\ba}{\begin{array}}
\newcommand{\ea}{\end{array}}

\newcommand{\pr}{\prime}

\begin{document}

\title{Visualizing the charge order and topological defects in an overdoped (Bi,Pb)$_2$Sr$_2$CuO$_{6+x}$ superconductor}
\author{Ying Fei}
\affiliation{Department of Physics, Zhejiang University, Hangzhou 310027, China}
\author{Yuan Zheng}
\email{phyyzheng@zju.edu.cn}
\affiliation{Department of Physics, Zhejiang University, Hangzhou 310027, China}
\author{Kunliang Bu}
\affiliation{Department of Physics, Zhejiang University, Hangzhou 310027, China}
\author{Wenhao Zhang}
\affiliation{Department of Physics, Zhejiang University, Hangzhou 310027, China}
\author{Ying Ding}
\affiliation{Beijing National Laboratory for Condensed Matter Physics, Institute of Physics, Academy of Sciences, Beijing 100190, China}
\author{Xingjiang Zhou}
\affiliation{Beijing National Laboratory for Condensed Matter Physics, Institute of Physics, Academy of Sciences, Beijing 100190, China}
\affiliation{Collaborative Innovation Center of Quantum Matter, Beijing 100871, China}
\author{Yi Yin}
\email{yiyin@zju.edu.cn}
\affiliation{Department of Physics, Zhejiang University, Hangzhou 310027, China}
\affiliation{Collaborative Innovation Center of Advanced Microstructures, Nanjing University, Nanjing 210093, China}


\begin{abstract}
Electronic charge order is a
symmetry breaking state in high-$T_\mathrm{c}$ cuprate superconductors. In scanning tunneling microscopy,
the detected charge-order-induced modulation is an electronic response of the charge order.
For an overdoped (Bi,Pb)$_2$Sr$_2$CuO$_{6+x}$ sample, we apply scanning tunneling microscopy to explore
local properties of the charge order.
The ordering wavevector is non-dispersive with energy, which can be confirmed and determined.
By extracting its order-parameter field, we identify dislocations
in the stripe structure of the electronic modulation, which correspond to topological defects
with an integer winding number of $\pm 1$. Through differential conductance maps over a
series of reduced energies, the development of different response of the charge order
is observed and a spatial evolution of topological defects is detected. The intensity of
charge-order-induced modulation increases with energy
and reaches its maximum when approaching the pseudogap energy. In this evolution,
the topological defects decrease in density and migrate in space.
Furthermore, we observe appearance and disappearance of closely spaced pairs of defects as energy changes.
Our experimental results could inspire further studies of the charge order in both
high-$T_\mathrm{c}$ cuprate superconductors and other charge density wave materials.

\end{abstract}

\maketitle

\section*{Introduction}
\label{sec1}
The electronic charge order, accompanying the pseudogap (PG) state, is an interesting phenomenon
in high-$T_\mathrm{c}$ cuprate superconductors~\cite{keimer2015quantum,timusk1999pseudogap,Vojta2009stripe}.
In the electronic phase diagram, the PG state and the charge order emerge when a
parent Mott insulator is doped with charge carriers~\cite{cai2016visualizing}.
With the increase of doping level,
superconductivity is developed and coexists with both the PG and charge order states over a broad doping range.
The study of the PG and charge order states may help to unravel the mechanism of
high-$T_\mathrm{c}$ superconductivity~\cite{Vojta2009stripe,cai2016visualizing,griiner1994density,hoffman2002four,
kohsaka2007intrinsic,vershinin2004local,parker2010silva,da2014ubiquitous,kohsaka2008cooper,wise2008,wise2009,
mesaros2011,fujita2014direct,comin2014charge,peng2017robust,peng2016direct,
comin2015symmetry,chang2012direct,wu2011magnetic,HamidianNatPhys2016,MesarosPNAS2016,fujita2014simultaneous,
lawler2010intra,webbArxiv}.

As a powerful tool of detecting the electronic structure with atomic resolution~\cite{fischer},
scanning tunneling microscopy (STM) has been extensively applied to investigate the charge order of cuprates
~\cite{cai2016visualizing,hoffman2002four,
kohsaka2007intrinsic,vershinin2004local,parker2010silva,da2014ubiquitous,kohsaka2008cooper,
wise2008,wise2009,mesaros2011,fujita2014direct,
fujita2014simultaneous,lawler2010intra,webbArxiv,HamidianNatPhys2016,MesarosPNAS2016}.
In an STM conductance map 
, the response of the charge order is usually identified by a checkerboard-like modulation
along two perpendicular CuO bond directions. In the Fourier-transformed data, the checkerboard-like modulation
is represented by four peaks centered at $\pm\boldsymbol{q}_x^\ast$ and $\pm\boldsymbol{q}_y^\ast$.
The absolute values of the charge order peaks
are inversely proportional to the periodicity $a_0/\delta$ of the real-space modulation,
given by $|\boldsymbol{q}_x^\ast|= |\boldsymbol{q}_y^\ast|=2\pi\delta/{a_0}$ with $a_0$ the lattice constant.
The checkerboard-like modulation with $\delta\approx3/4$ has been well studied~\cite{kohsaka2007intrinsic,parker2010silva,kohsaka2008cooper,mesaros2011,fujita2014simultaneous,
lawler2010intra}, and is prominent at high energies around the PG magnitude.
Recently, the charge order of $\delta\approx3/4$ has been discovered to be
an electronic modulation with a $d$-form factor~\cite{fujita2014direct,HamidianNatPhys2016,MesarosPNAS2016},
compatible with the charge order of $\delta\approx1/4$
detected with different experimental techniques~\cite{Vojta2009stripe,cai2016visualizing,hoffman2002four,vershinin2004local,da2014ubiquitous,
wise2008,wise2009,comin2014charge,peng2017robust,peng2016direct,comin2015symmetry,webbArxiv}.

The STM studies of the charge order are mainly implemented in two families of cuprates, Bi$_2$Sr$_2$CaCu$_2$O$_{8+x}$ (Bi-2212)
and Bi$_2$Sr$_2$CuO$_{6+x}$ (Bi-2201). With respect to the crystal structure, two CuO layers exist
in a unit cell of Bi-2212, while only one in that of Bi-2201. The electronic structures of these two cuprates
are different as well. The PG state vanishes in the middle of the superconducting regime in
Bi-2212~\cite{fujita2014simultaneous},
while it extends to the overdoped regime in Bi-2201~\cite{heyang2014,zheng2017the}.
Therefore the charge order in Bi-2201 can be explored in a broad range of doping. The determination of
the ordering wavevector has been a main focus in previous studies~\cite{MesarosPNAS2016,webbArxiv}. With the increase of doping, a
commensurate to incommensurate transition of the wavevector has been discovered~\cite{webbArxiv}.

The structural disorder in doped cuprates can induce spatial fluctuations in the electronic orders.
In the Ginzburg-Landau theory, the charge order can be described by an order-parameter field $\psi(\boldsymbol{r})=A(\boldsymbol{r})\exp[i\phi(\boldsymbol{r})]$,
where the amplitude $A(\boldsymbol{r})$ and the phase $\phi(\boldsymbol{r})$ vary spatially. In general,
charge order is a long-range order with local fluctuations resulting from perturbations.
Furthermore, dislocations may interrupt the long range order, with dislocation cores represent
singularities in the order parameter phase $\phi(\boldsymbol{r})$. A nonzero integer winding number
is obtained along any path enclosing such a singular point, thus named a topological defect.
Previous work examining underdoped Bi-2212 extracted the spatial structure of local fluctuations in the
charge-order-induced modulation, revealing a close connection to nematicity~\cite{mesaros2011}. A similar study examining the
local structure of the charge-order-induced signal in Bi-2201 is a natural extension.

In this paper, we apply STM to study the charge order in overdoped Bi-2201.
A non-dispersive and incommensurate wavevector is determined to be at $\delta\approx0.83$.
From the local amplitude and phase of the modulation, we identify singular topological defects
in the order parameter field, corresponding to dislocation cores in the filtered stripe structure
of the charge-order-induced modulation. Around a single defect, the data analysis leads to a generic
phase slip picture. As energy increases away from the Fermi level,
the response of the charge order is gradually enhanced, and the topological defects migrate in space.
The appearance and disappearance of defect pairs are also observed.
The defect `movement' can be related to the transfer of dislocation cores in regions of bent stripes.
The total number of defects decreases with energy until it saturates, which implies that the intensity of
modulation can affect how robust the response of charge order is against local perturbations.
This energy-dependent phenomenon can be further investigated in other cuprate superconductors and
charge density wave materials.

\section*{Materials and method}
\label{sec2}
The high-quality (Bi,Pb)$_2$Sr$_2$CuO$_{6+x}$  single crystal in our experiment is grown by the
traveling-solvent floating-zone method~\cite{lin2010high}. With a size of $2$ mm $\times$ 2 mm $\times$ 0.1 mm,
the sample is cut from an as-grown ingot, followed  by being post-annealed at a specific temperature and under a controlled atmosphere.
The critical temperature of the Pb-doped Bi-2201 sample studied in this work is $T_c\approx 13$ K.
The hole doping for this overdoped (OD13K) sample is estimated to be $p\approx0.21$ according to the Ando formula~\cite{andoformula,FeiyingSCPMA}.

All the data in this paper are taken in an ultra-high-vacuum STM~\cite{zheng2017the}. The Bi-2201 sample is cleaved at liquid-nitrogen temperature
and immediately inserted into the STM head. The measurement is performed at liquid-helium temperature ($T\approx 4.5$ K).
The STM topography is taken at a sample bias of $V_\mathrm{b}=100$ mV and a setpoint current of $I_\mathrm{s}=100$ pA. The local
differential conductance ($dI/dV$) spectra are acquired simultaneously with the topography
by a standard lock-in technique with a modulation frequency of $f=983.4$ Hz.

\section*{Results}
\label{sec3}
\begin{figure}
\includegraphics[width=0.85\columnwidth]{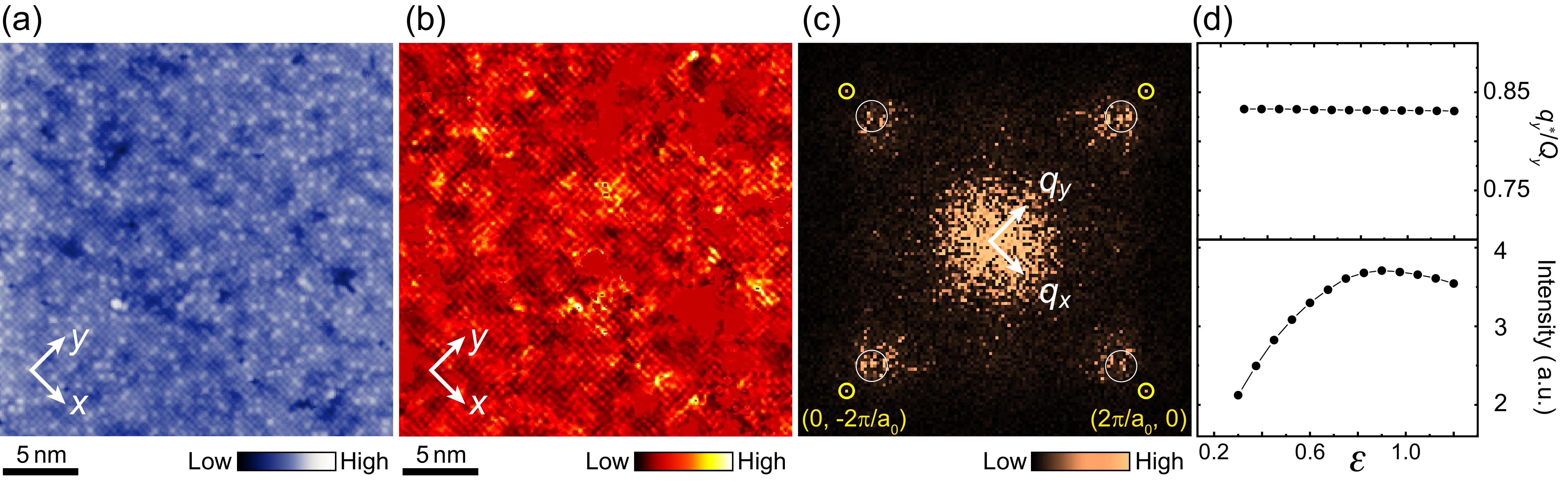}
\caption{Topography and electronic properties of an overdoped Bi-2201 sample.
         ({\bf a}) A 27 nm $\times$ 27 nm topographic image taken on a cleaved BiO surface.
        ({\bf b}) The ratio map $Z(\boldsymbol{r},\varepsilon=1.05)$ in the same FOV as in panel ({\bf a}), with the VHS regions filled in the averaged value.
         ({\bf c}) The Fourier transform of panel ({\bf b}). The four Bragg peaks
         are highlighted by yellow circles, while the charge order peaks
         are highlighted by white circles.
         ({\bf d}) In the top and bottom panels, the position and intensity of the charge order peak are extracted as a function of the reduced energy $\varepsilon$, respectively.
         }
\label{fig:1}
\end{figure}

\textbf{Topography and charge order of overdoped Bi-2201.}
We work with a high-quality overdoped (Bi,Pb)$_2$Sr$_2$CuO$_{6+x}$ single crystal sample~\cite{lin2010high,zheng2017the} whose critical
temperature $T_c\approx 13$ K and hole doping $p\approx0.21$~\cite{andoformula,FeiyingSCPMA}. The data is taken
in an ultra-high-vacuum STM at $T \approx 4.5$ K~\cite{zheng2017the}. Figure~\ref{fig:1}(a) is a topographic
image of 27 nm $\times$ 27 nm  taken on a cleaved BiO surface,
showing a square lattice of Bi atoms with the lattice constant $a_0\approx3.8$ {\AA}.
In the CuO layer, each Cu atom is located directly below a Bi atom, sharing the same lattice constant~\cite{HamidianNJP2012}.
The brighter spots in Fig.~\ref{fig:1}(a)
are Pb atoms, which are partially substituted for Bi atoms and strongly suppress
the structural supermodulation.

The differential conductance spectra
are measured by varying the voltage $V$, in which the PG magnitudes ($\Delta_\mathrm{PG}$)
can be extracted from coherence-peak positions.
The local electronic structure is probed by a differential conductance map,
$g(\boldsymbol{r},E=eV)=dI/dV(\boldsymbol{r},V)$.
In our Pb-doped Bi-2201 sample, the distribution of both the differential conductance and the PG magnitude
is spatially inhomogeneous (see Supplementary I).
To exclude the spatial inhomogeneity of the PG magnitude, the conductance map is rescaled
as $g(\boldsymbol{r},\varepsilon)$ with a reduced energy $\varepsilon=E/\Delta_{\mathrm{PG}}(\boldsymbol{r})$.
Subsequently, we calculate a ratio map, $Z(\boldsymbol{r},\varepsilon)=g(\boldsymbol{r},\varepsilon)/g(\boldsymbol{r},-\varepsilon)$,
which reduces the systematic error known as the setpoint effect in STM measurement~\cite{kohsaka2007intrinsic,kohsaka2008cooper,heyang2014}.
In this field of view (FOV), we also find some zero gap patches
which can be attributed to the van Hove singularity (VHS) states~\cite{fischer,zheng2017the,yayuNJP}. A spatially averaged
value of $Z(\boldsymbol{r},\varepsilon)$ is applied to fill in the VHS regions~\cite{zheng2017the}.
As the differential conductance is proportional to the electronic density of states, the response
of charge order is reflected in the $Z(\boldsymbol{r}, \varepsilon)$-maps at different
reduced energies, which is the main focus of data analysis in this paper.

Figure~\ref{fig:1}(b) displays a ratio map, $Z(\boldsymbol{r},\varepsilon=1.05)$, at the energy scale of the PG and
in the same FOV as in Fig.~\ref{fig:1}(a). A small deviation of $\varepsilon$ from unity
is due to our technique of binning the reduced energies.
A checkerboard-like spatial modulation is observed, which represents a response of the static charge order in recorded STM results.
To maintain an atomic registry across the FOV, drift correction of the lattice is applied to both topographic and electronic data~\cite{lawler2010intra}.
To quantify the spatial periodicity of charge-order-induced modulation at the PG energy,
the Fourier transform of Fig.~\ref{fig:1}(b), $\tilde{Z}(\boldsymbol{q}, \varepsilon)=\mathrm{FT}[Z(\boldsymbol{r},\varepsilon)]$,
is shown in the momentum $\boldsymbol{q}$-space in Fig.~\ref{fig:1}(c).
Four sharp Bragg peaks located at $\pm\boldsymbol{Q}_x$ and $\pm\boldsymbol{Q}_y$ are observed,
each collapsed into a single pixel due to the lattice drift correction~\cite{zheng2017the,lawler2010intra}.
The Bragg wavevectors, $\boldsymbol{Q}_x=(2\pi/{a_0},0)$ and
 $\boldsymbol{Q}_y=(0, 2\pi/{a_0})$,  are consistent with the square lattice structure of Bi atoms.
In addition, there exist four peaks near the Bragg peaks,
which correspond to the checkerboard-like spatial modulation induced by the charge order.
The centers of the four charge order peaks are estimated to be at $\pm\boldsymbol{q}^\ast_x\approx\pm0.83\boldsymbol{Q}_x$
and $\pm\boldsymbol{q}^\ast_y\approx\pm0.83\boldsymbol{Q}_y$ from Gaussian fitting (Supplementary II).
The same analysis can be applied to $\tilde{Z}(\boldsymbol{q},\varepsilon)$-maps at different reduced energies.
In the top panel of Fig.~\ref{fig:1}(d), we plot the amplitude $|\boldsymbol{q}^\ast_{x/y}|$ of the charge-order wavevectors
as a function of reduced energy $\varepsilon$. As $\varepsilon$ changes, the wavevectors $\boldsymbol{q}^\ast_{x/y}$ are nearly
invariant, suggesting a non-dispersive and static charge order. The wavevectors are incommensurate with the lattice, consistent with the result
for overdoped Bi-2201 in Ref.~\cite{webbArxiv}. In the bottom panel of Fig.~\ref{fig:1}(d),
we also plot the intensity of the charge-order peak as a
function of $\varepsilon$, showing that the response of the charge order
is weak at small values of $\varepsilon$ and becomes intensified around the PG energy.
A more detailed discussion of the $\varepsilon$-dependence will be provided later.

\begin{figure}
\includegraphics[width=0.75\columnwidth]{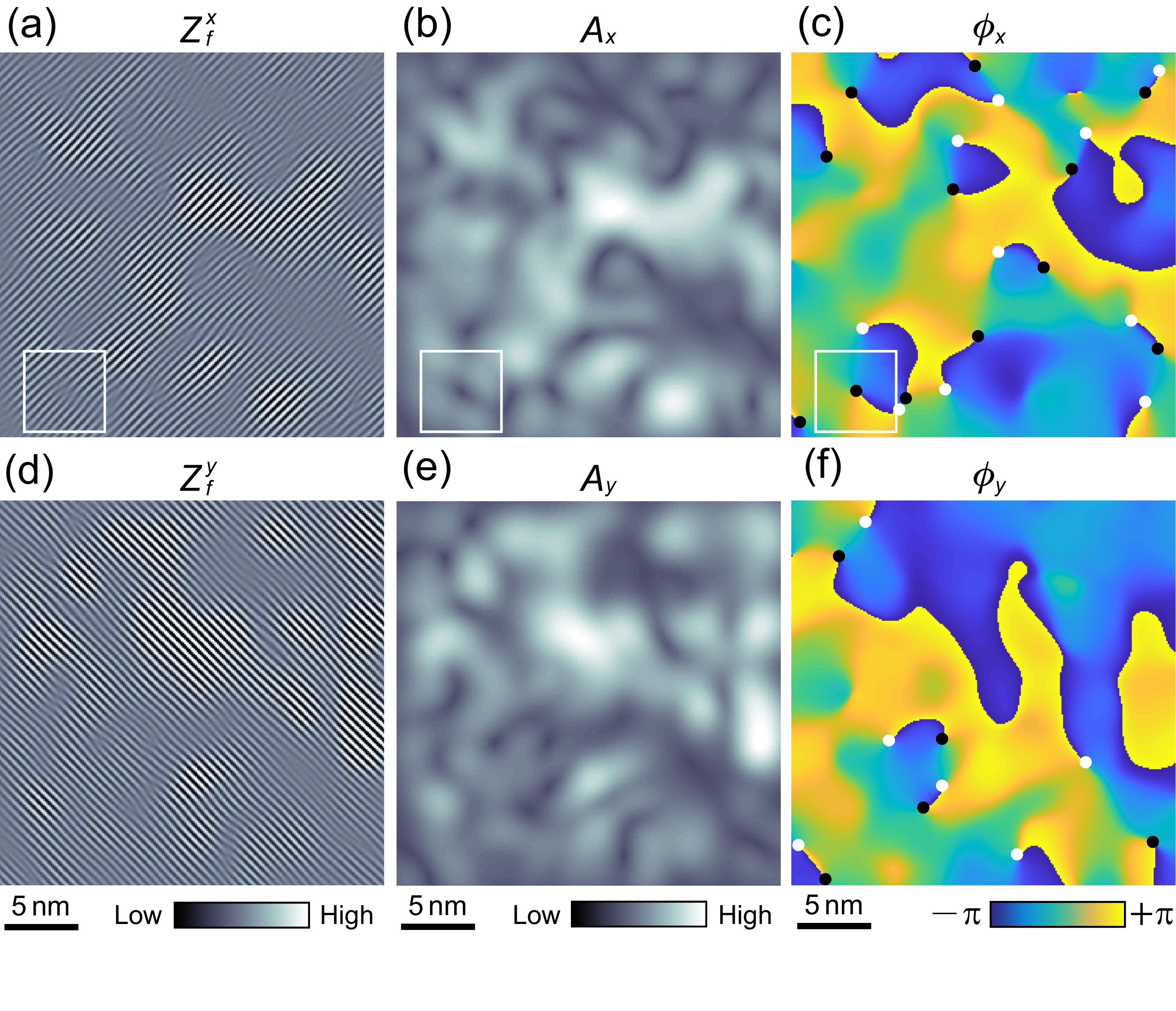}
\caption{Maps of the charge order after a Gaussian filtering. ({\bf a}) $Z^x_f(\boldsymbol{r},\varepsilon=1.05)$,
          ({\bf b}) $A_x(\boldsymbol{r},\varepsilon=1.05)$, and ({\bf c}) $\phi_x(\boldsymbol{r},\varepsilon=1.05)$
          describe $x$-direction modulations.
          ({\bf d}) $Z^y_f(\boldsymbol{r},\varepsilon=1.05)$, ({\bf e}) $A_y(\boldsymbol{r},\varepsilon=1.05)$, and ({\bf f}) $\phi_y(\boldsymbol{r},\varepsilon=1.05)$ describe $y$-direction modulations. In ({\bf c}) and ({\bf f}),
          topological defects with positive and negative polarities are shown with white and black circles, respectively.
 }
\label{fig:2}
\end{figure}

\vspace{12pt}
\textbf{Identification of topologaical defects.}
To explore the spatial variation of the checkerboard-like modulation, we first separate the $x$- and $y$-components
with a Gaussian filtering technique. The $x$-component of the modulation in the momentum space is extracted as
\be
\tilde{Z}^x_f(\boldsymbol{q},\varepsilon)=\tilde{Z}(\boldsymbol{q},\varepsilon)[\tilde{f}(\boldsymbol{q}+\boldsymbol{q}^\ast_x)+\tilde{f}(\boldsymbol{q}-\boldsymbol{q}^\ast_x)],
\ee
where $\tilde{f}(\boldsymbol{q})=\exp(-q^2\Lambda^2/2)$ is a Gaussian filtering function.
In practice, the cutoff size is chosen at $\Lambda=1.2$ nm.
Following an inverse Fourier transform, $Z^x_f(\boldsymbol{r},\varepsilon)=\mathrm{FT}^{-1}[\tilde{Z}^x_f(\boldsymbol{q},\varepsilon)]$,
we obtain the checkerboard-like modulation along the $x$-direction in the real space. The same technique is applied to extract
the $y$-components, $\tilde{Z}^y_f(\boldsymbol{q},\varepsilon)$ and $Z^y_f(\boldsymbol{r},\varepsilon)$.
The final results of $Z^x_f(\boldsymbol{r},\varepsilon=1.05)$ and $Z^y_f(\boldsymbol{r},\varepsilon=1.05)$ at the PG energy are
plotted in Figs.~\ref{fig:2}(a) and~\ref{fig:2}(d), from which we observe stripes along the $y$-
and $x$-directions, respectively. In general, the $x$-component of the charge order can be re-expressed as
\be
Z^x_f(\boldsymbol{r},\varepsilon)=A_x(\boldsymbol{r}, \varepsilon)\cos[\boldsymbol{q}_x^\ast\cdot \boldsymbol{r}+\phi_x(\boldsymbol{r}, \varepsilon)],
\label{eq_002}
\ee
where $A_x(\boldsymbol{r}, \varepsilon)$ and $\phi_x(\boldsymbol{r}, \varepsilon)$ are
the amplitude and phase of the order-parameter field at location $\boldsymbol{r}$.
Here another Gaussian filtering process is applied to extract $\phi_x(\boldsymbol{r}, \varepsilon)$ around the charge order wavevector~\cite{mesaros2011}, given by
\be
\tan \phi_x(\boldsymbol{r}, \varepsilon) = \frac{\int d\boldsymbol{r}^\pr Z^x_f(\boldsymbol{r}^\pr,\varepsilon)\sin [\boldsymbol{q}_x^\ast\cdot \boldsymbol{r}^\pr]f(\boldsymbol{r}-\boldsymbol{r}^\pr)}
{\int d\boldsymbol{r}^\pr Z^x_f(\boldsymbol{r}^\pr,\varepsilon)\cos [\boldsymbol{q}_x^\ast\cdot \boldsymbol{r}^\pr]f(\boldsymbol{r}-\boldsymbol{r}^\pr)},
\ee
with $f(\boldsymbol{r})=\mathrm{FT}^{-1}[\tilde{f}(\boldsymbol{q})]$. The amplitude $A_x(\boldsymbol{r}, \varepsilon)$ is then calculated using
the definition in Eq.~(\ref{eq_002}). Accordingly, the order-parameter field associated with the $x$-component of the charge order
is given by $\psi_x(\boldsymbol{r})=A_x(\boldsymbol{r})\exp[i\phi_x(\boldsymbol{r})]$.
In Figs.~\ref{fig:2}(b) and~\ref{fig:2}(c), we show $A_x(\boldsymbol{r}, \varepsilon=1.05)$
and $\phi_x(\boldsymbol{r}, \varepsilon=1.05)$ extracted from Fig.~\ref{fig:2}(a).
The amplitude and phase are inhomogeneous, unlike the uniform order parameter of ideal long range order.
The three spatially resolved quantities, $Z^x_f(\boldsymbol{r},\varepsilon=1.05)$, $A_x(\boldsymbol{r}, \varepsilon=1.05)$
and $\phi_x(\boldsymbol{r}, \varepsilon=1.05)$ in Figs.~\ref{fig:2}(a)-\ref{fig:2}(c), display correlated behavior.
The same approach is applied to the $y$-component of the charge-order-induced modulation ($Z^y_f(\boldsymbol{r},\varepsilon=1.05)$ in Fig.~\ref{fig:2}(d)),
and the resulting $A_y(\boldsymbol{r}, \varepsilon=1.05)$ and $\phi_y(\boldsymbol{r}, \varepsilon=1.05)$
are plotted in Figs.~\ref{fig:2}(e) and~\ref{fig:2}(f), respectively.

\begin{figure}
\includegraphics[width=0.75\columnwidth]{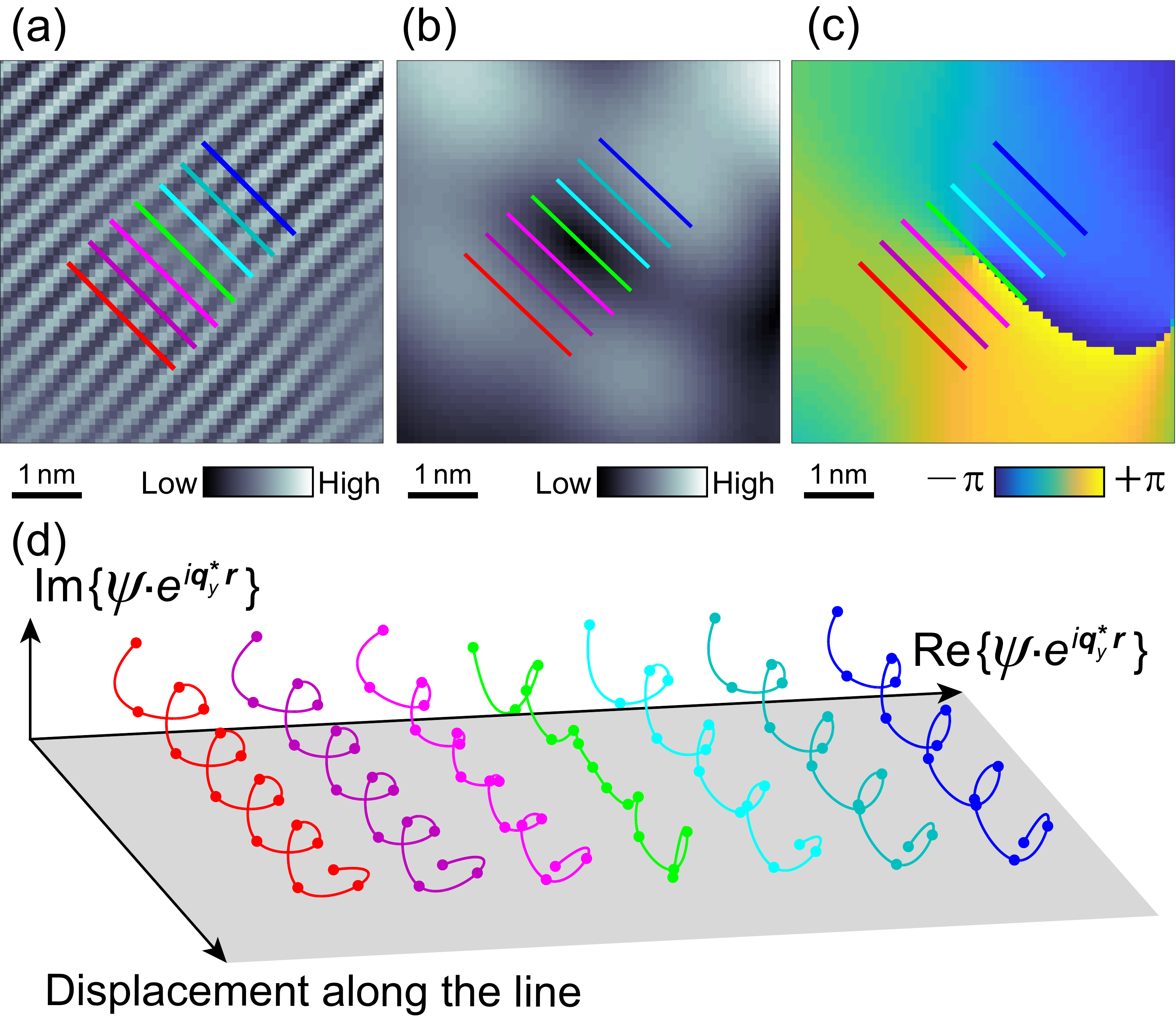}
\caption{Enlarged maps of ({\bf a}) $Z^x_f(\boldsymbol{r},\varepsilon=1.05)$,
          ({\bf b}) $A_x(\boldsymbol{r},\varepsilon=1.05)$, ({\bf c}) $\phi_x(\boldsymbol{r},\varepsilon=1.05)$ around
          a selected topological defect in the solid box in Figs.~\ref{fig:2}(a)-\ref{fig:2}(c). In ({\bf a})-({\bf c}), seven lines
         perpendicular to the stripe direction are chosen to demonstrate the phase slip picture.
         ({\bf d}) Along the seven chosen linecuts in ({\bf a})-({\bf c}), the real and imaginary parts of $\psi_x(\boldsymbol{r})\exp[i\boldsymbol{q}_x^\ast\cdot\boldsymbol{r}]$
         are displayed as a function of the displacement along each linecut. The experimentally extracted data are shown in solid dots
         and the spirals (solid lines) are obtained by the cubic interpolation method.
}
\label{fig:3}
\end{figure}

From Figs.~\ref{fig:2}(a) and~\ref{fig:2}(d), we observe that the stripes of charge-order-induced modulation
are distorted in space, including weak fluctuations, as well as strong disruptions.
As an illustration, we select a small area in the solid box in Fig.~\ref{fig:2}(a) and
plot magnified maps of $Z^x_f(\boldsymbol{r},\varepsilon=1.05)$, $A_x(\boldsymbol{r},\varepsilon=1.05)$ and $\phi_x(\boldsymbol{r},\varepsilon=1.05)$ in Figs.~\ref{fig:3}(a)-\ref{fig:3}(c).
Near the middle of the image, a single stripe splits into two stripes along the $y$-direction, analogous to an
edge dislocation in a crystal lattice. The manifestation of the branch point, or dislocation core, in
$A_x(\boldsymbol{r})$ and $\phi_x(\boldsymbol{r})$ is associated with the presence of a singularity:
(1) The phase $\phi_x(\boldsymbol{r})$ is undefined at the dislocation core,
while a winding phase of $2\pi$ is acquired for $\phi_x(\boldsymbol{r})$ if a clockwise cycle
is taken around this singularity. (2) To generate a physically meaningful order parameter, the
corresponding amplitude $A_x(\boldsymbol{r})$ is suppressed to zero, as shown by a dark region near the middle of Fig.~\ref{fig:3}(b).
To further visualize the behavior of the singularity,
seven linecuts along the $y$-direction are
chosen around the dislocation in Fig.~\ref{fig:3}(a). For each linecut, the real and imaginary parts of
$\psi_x(\boldsymbol{r})\exp[i\boldsymbol{q}_x^\ast\cdot\boldsymbol{r}]$ are plotted
as a function of the spatial location $\boldsymbol{r}$ (see Fig.~\ref{fig:3}(d)). Along the right top linecut
(in blue color), a spiral of $\psi_x(\boldsymbol{r})\exp[i\boldsymbol{q}_x^\ast\cdot\boldsymbol{r}]$
propagates with a roughly constant amplitude and four turns are generated within the selected distance.
As the linecut approaches the dislocation core (the green line), the spiral is disrupted and the middle turn is
broken into two very small turns. An extra phase of $2\pi$ is acquired through $\phi_x(\boldsymbol{r})$
so that the number of the spiral turns is increased from 4 to 5 along the linecuts below the green one,
similar to the phase slip picture for other ordered states~\cite{chaikin1995principles,tinkham1996introduction}.
Through a combined analysis of $Z^{x/y}_f(\boldsymbol{r},\varepsilon)$, $A_{x/y}(\boldsymbol{r},\varepsilon)$ and $\phi_{x/y}(\boldsymbol{r},\varepsilon)$,
the dislocation cores behave as topological defects in the charge-order-induced modulation,
and almost do not affect the wavevector of the modulation (see Fig.~S3 in Supplementary III).

With $\phi_{x/y}(\boldsymbol{r})$ defined modulo $2\pi$, a branch cut following a curved line
can be emanating from each defect (see Figs.~\ref{fig:2}(c) and~\ref{fig:2}(f)). Crossing
a branch cut generates an artificial jump of $\pm2\pi$, which however does not affect the value of the order parameter $\psi_{x/y}(\boldsymbol{r})$.
In this FOV, most of these branch cuts connect two defects with opposite polarities defined by the signs of their winding phases.
In general, topological defects are emergent excitations in an ordered state, such as quantized vortices in
superfluids and superconductors~\cite{chaikin1995principles,tinkham1996introduction}. Compared to a single unbounded defect,
a pair of defects with opposite polarities have a lower excitation energy associated with an attractive interaction.
In our system, defect pairing is observed in Figs.~\ref{fig:2}(c) and~\ref{fig:2}(f), except for those around the corners of the FOV.

\vspace{12pt}
\textbf{Evolution of the charge-order-induced modulation and topological defects.}
In cuprates such as Bi-2201 superconductors, various electronic orders coexist and can compete with each other.
In Fig.~\ref{fig:1}(d), we find that response of the charge order appears at a low $\varepsilon$ and is intensified at
the PG energy, consistent with a consensus that the charge order is correlated with the PG state~\cite{parker2010silva,fujita2014simultaneous,vershinin2004local,wise2009,comin2014charge,heyang2014}.
Next we explore the spatial variation of the charge-order-induced modulation at different reduced energies.
The above analysis procedure is applied to $Z(\boldsymbol{r}, \varepsilon)$ maps of different $\varepsilon$ to obtain
the $x$- and $y$-components of the modulation, as well as their amplitudes and phases. From a series of the filtered maps of
$Z^x_f(\boldsymbol{r}, \varepsilon)$ and $Z^y_f(\boldsymbol{r}, \varepsilon)$ (see Supplementary Fig. S6 and S8),
we observe that the stripe structure of the modulation is maintained
when $\varepsilon$ increases. However, we also find visible changes at certain regions.
The one-dimensional structure of a stripe is easily bent by local stress.
Neighboring stripes are bent similarly but with gradually decaying strengths.
Under large stress, dislocations can be created among the bent stripes.
As $\varepsilon$ changes, this local stress may be released or enhanced, causing
`movement' of defects and even their appearance and disappearance. We emphasize that
the charge order itself is a static electronic order without energy dependence. The
$\varepsilon$-dependent phenomenon discussed here is the different
response of the charge order represented in the STM spectroscopy measurement.

\begin{figure}
\includegraphics[width=0.75\columnwidth]{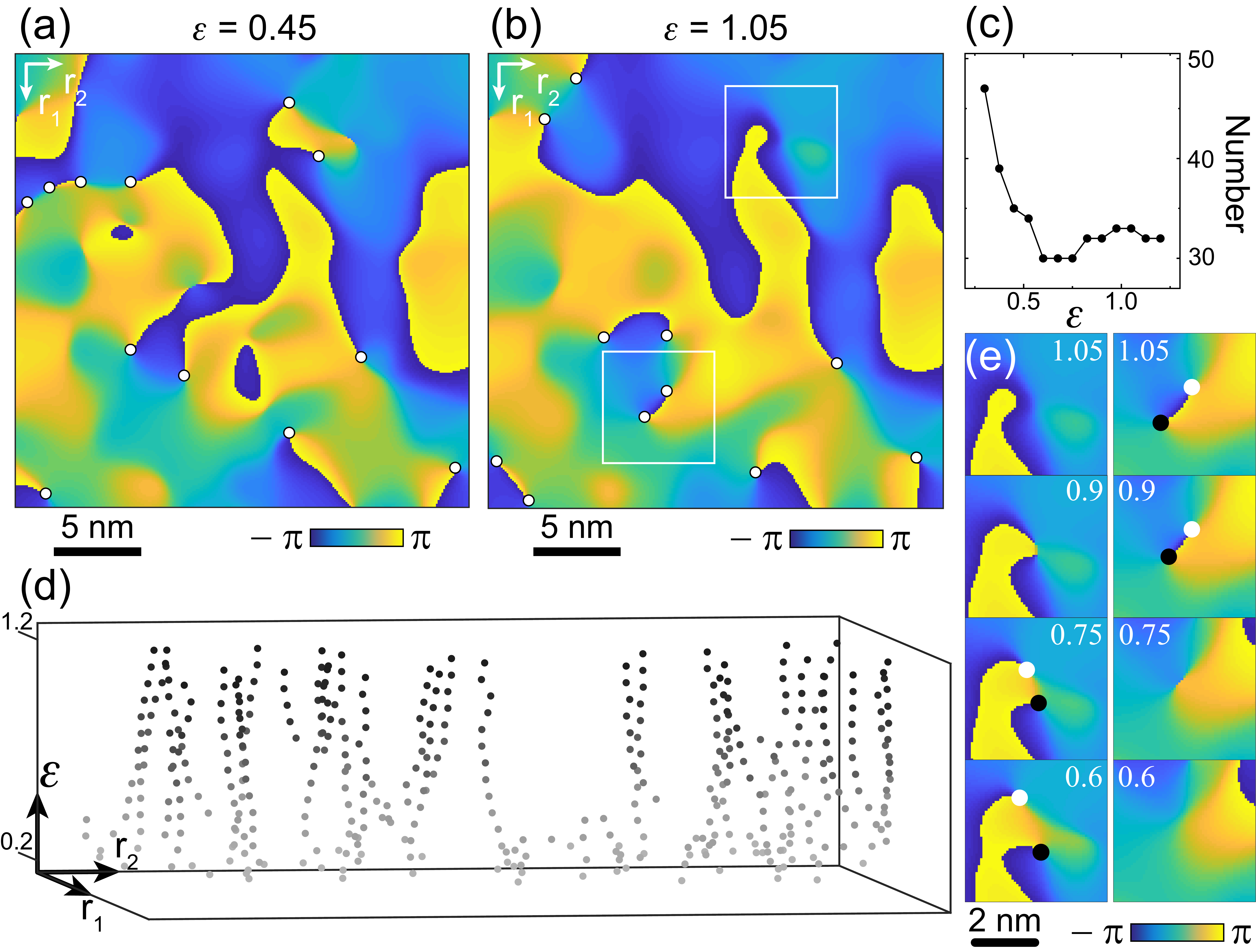}
\caption{Reduced-energy dependence of the distribution of topological defects.
          ({\bf a}) and ({\bf b}) are the two phase maps, $\phi_x(\boldsymbol{r},\varepsilon=0.45)$ and $\phi_x(\boldsymbol{r},\varepsilon=1.05)$, respectively.
          Topological defects are marked by white circles in both maps.
          ({\bf c}) The total number of defects as a function of the reduced energy.
          ({\bf d}) The spatial distribution of defects as a function of the reduced energy.
          Here $\boldsymbol{r}_1$ and $\boldsymbol{r}_2$ represent two perpendicular axes of the FOV.
          ({\bf e}) Enlarged phase maps of showing how a pair of defects appear (right column) or disappear (left column)
          when the reduced energy increases. The area of these maps are shown by white boxes in ({\bf b}).
          }
\label{fig:4}
\end{figure}

To visualize the $\varepsilon$-dependence of the detailed stripe structure,
we inspect the phase maps of $\phi_{x/y}(\boldsymbol{r}, \varepsilon)$.
As shown in Figs.~\ref{fig:4}(a) and \ref{fig:4}(b), the
overall structure of two phase maps, $\phi_x(\boldsymbol{r}, \varepsilon=0.45)$ and $\phi_x(\boldsymbol{r}, \varepsilon=1.05)$,
are similar to each other, demonstrating the stability of the charge-order-induced modulation but with subtle changes.
As shown in Fig.~\ref{fig:4}(c),
the total number of defects, extracted from both $\phi_x(\boldsymbol{r}, \varepsilon)$ and $\phi_y(\boldsymbol{r}, \varepsilon)$, gradually decreases
with $\varepsilon$ and is stabilized around 30 when $\varepsilon$ is larger than 0.5.
We record the spatial locations of defects at different $\varepsilon$ and present a
three-dimensional map in Fig.~\ref{fig:4}(d), showing the movement of defects.
The appearance and disappearance of defect pairs are observed,
and can even occur around the PG energy (see Supplementary VI for more details).
Two examples are shown in Fig.~\ref{fig:4}(e). In the left column, the distance between two defects gradually decreases
with the increasing $\varepsilon$. For $\varepsilon=0.9$ and 1.05, the two defects cannot be distinguished from each other 
because their distance is smaller than our cutoff size, demonstrating `annihilation' of a defect pair with opposite polarities.
The right column shows the `creation' of a defect pair connected by a branch cut.
For charge-order-induced modulation with a relatively low intensity,
the stripe bending is prone to induce pairs of closely spaced defects, leading
to the increasing number of defects at small $\varepsilon$. The `movement' of
defects corresponds to transfer of dislocation cores within a region of bent
stripes (see Supplementary Fig. S10).
Furthermore, we find a negligible correlation between the topological defects and inhomogeneous PG
(see Supplementary IV). Although the spatial resolution of detecting paired defects changes with the
cutoff size of the filtering function (or the coarse grained length equivalently),
the evolution of defects qualitatively holds for various cutoff sizes (see Supplementary V).

\section*{Conclusions}
\label{sec4}

In this paper, we investigate the charge order in an overdoped (Bi,Pb)$_2$Sr$_2$CuO$_{6+x}$ (Bi-2201)
sample with STM. In real space, the charge-order-induced modulation is identified in the
ratio $Z$-map of $Z(\boldsymbol{r}, \varepsilon)$. In momentum space, the modulation
is represented by four peaks around incommensurate wavevectors $\pm\boldsymbol{q}^\ast_x\approx\pm0.83(2\pi/{a_0},0)$ and $\pm\boldsymbol{q}^\ast_y\approx\pm0.83(0,2\pi/{a_0})$.
The charge-order-induced modulation starts to appear at low reduced energies.
With increasing $\varepsilon$, the modulation is gradually intensified and reaches its maximum strength around the PG energy.
The incommensurate wavevector is not dispersive with $\varepsilon$, consistent with the static charge order or charge density
wave order. The spatially-varied order-parameter fields enable identification of singular points,
i.e. topological defects. In consecutive order-parameter maps with changing $\varepsilon$, the positions of defects
gradually change, and pairs of defects appear and disappear. As energy approaches the PG energy, the number of defects
decreases and saturates. This phenomenon uncovers a new aspect of the charge-order response in STM measurement.
We expect to investigate whether the observed behavior is generic in cuprate superconductors or other charge
density wave materials, which requires future experimental and theoretical efforts.

\section*{Acknowledgements}
This work is supported by the National Basic Research Program
of China (2015CB921004), the National Natural Science Foundation of
China (NSFC-11374260), and the Fundamental Research Funds
for the Central Universities in China. XJZ thanks financial support from the National Natural Science Foundation of
China (NSFC-11334010), the National Key Research and Development Program of China (2016YFA0300300),
and the Strategic Priority Research Program (B) of the Chinese Academy of Sciences (XDB07020300).

%
%

\newpage

\end{document}


\title{Supplementary Information for \\ ``Visualizing the charge order and topological defects in an overdoped (Bi,Pb)$_2$Sr$_2$CuO$_{6+x}$ superconductor ''}

\author{Ying Fei}
\affiliation{Department of Physics, Zhejiang University, Hangzhou 310027, China}
\author{Yuan Zheng}
\email{phyyzheng@zju.edu.cn}
\affiliation{Department of Physics, Zhejiang University, Hangzhou 310027, China}
\author{Kunliang Bu}
\affiliation{Department of Physics, Zhejiang University, Hangzhou 310027, China}
\author{Wenhao Zhang}
\affiliation{Department of Physics, Zhejiang University, Hangzhou 310027, China}
\author{Ying Ding}
\affiliation{Beijing National Laboratory for Condensed Matter Physics, Institute of Physics, Academy of Sciences, Beijing 100190, China}
\author{Xingjiang Zhou}
\affiliation{Beijing National Laboratory for Condensed Matter Physics, Institute of Physics, Academy of Sciences, Beijing 100190, China}
\affiliation{Collaborative Innovation Center of Quantum Matter, Beijing 100871, China}
\author{Yi Yin}
\email{yiyin@zju.edu.cn}
\affiliation{Department of Physics, Zhejiang University, Hangzhou 310027, China}
\affiliation{Collaborative Innovation Center of Advanced Microstructures, Nanjing University, Nanjing 210093, China}


\setcounter{figure}{0}
\setcounter{equation}{0}
\setcounter{section}{0}
\makeatletter
\renewcommand{\thefigure}{S\@arabic\c@figure}

\maketitle

\begin{figure}[tp]
\centering
\includegraphics[width=1.1\columnwidth]{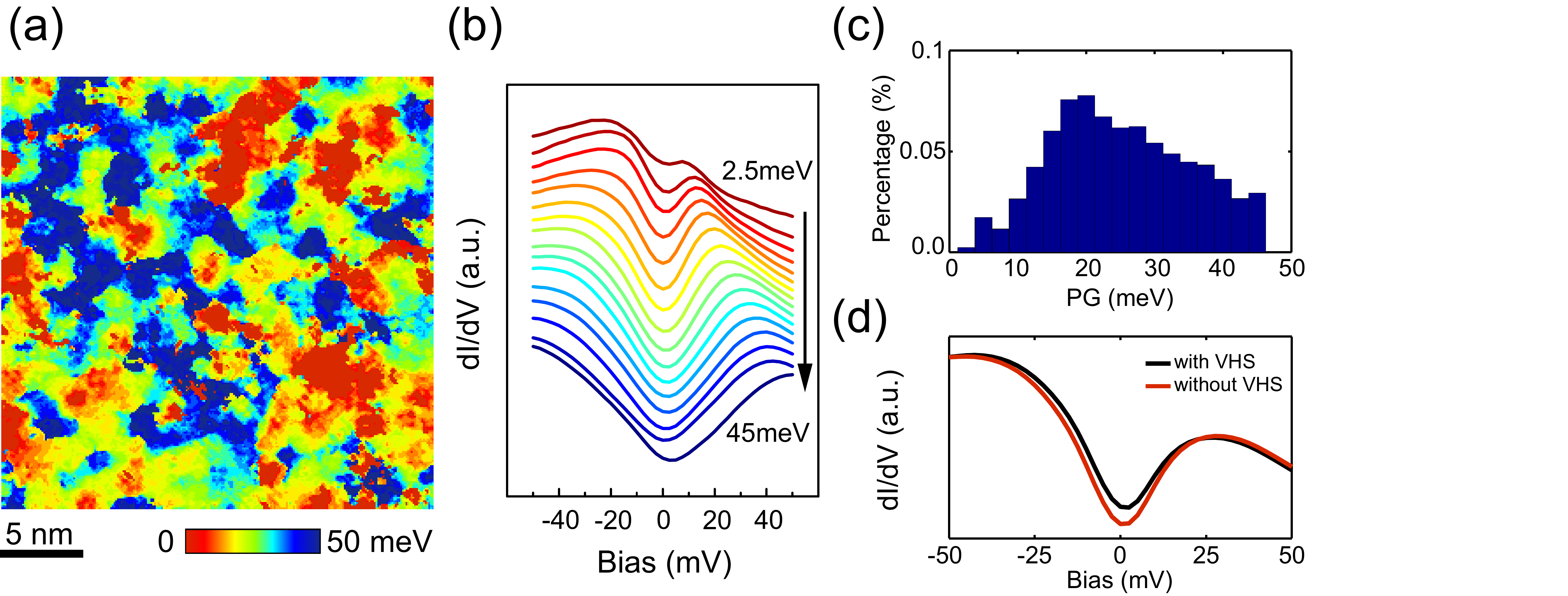}
\caption{(a) A 27 nm $\times$ 27 nm pesudogap map in the same field of view as in Fig.~1(a) in the main text.
(b) The average $dI/dV$ spectra over a bin size of $\delta\Delta_\mathrm{PG}=2.5$ meV with
respect to PG magnitude $\Delta_\mathrm{PG}$ ranging from 2.5 meV to 45 meV.
(c) Histogram of the PG magnitudes.
(d) The average $dI/dV$ spectra with and without the van Hove singularity states.
}
\label{figS1}
\end{figure}

\section{Spatially inhomogeneous differential conductance spectra }
\label{secs1}

The sample studied in this paper is an overdoped (OD) (Bi,Pb)$_2$Sr$_2$CuO$_{6+x}$ (Bi-2201)
single crystal with the superconducting critical temperature of $T_c\approx 13$ K and the hole doping of $p\approx 0.21$.
For the field of view (FOV) shown in Fig.~1(a) of the main text, a series of $dI/dV$ spectra have been
taken simultaneously with the topographic image at a dense array of locations.
The pseudogap (PG) magnitude ($\Delta_\mathrm{PG}$) at each location
is determined by extracting the bias voltage of the positive coherence
peak. The symmetric negative coherence peaks are hidden in the filled-state spectra~\cite{boyer, caipeng, fischer}.
The resulting PG map is shown in Fig.~\ref{figS1}(a), from which a nanoscale
inhomogeneity can be observed. This inhomogeneity is also illustrated in
spatially averaged spectra binned by the PG magnitudes (see Fig.~\ref{figS1}(b)).
The broad distribution of PG, varying from 2.5 meV to 45 meV,
is represented by a histogram in Fig.~\ref{figS1}(c).
With a peak value at $\Delta_\mathrm{PG}\approx19$ meV, the distribution of PG in this overdoped Bi-2201 is relatively broader than the one in another
overdoped Bi-2201 with a similar $T_\mathrm{c}$ of 15 K~\cite{heyangScience}.
In this FOV, we also find around $10.9\%$
area with zero gap, which is attributed to the van Hove singularity (VHS) state~\cite{ZY,fischer}. Figure~\ref{figS1}(d)
presents the average spectra with and without van Hove singularity states. The two spectra are similar
to each other, except for a small difference of the zero bias conductance.

\begin{figure}[tp]
\centering
\includegraphics[width=0.90\columnwidth]{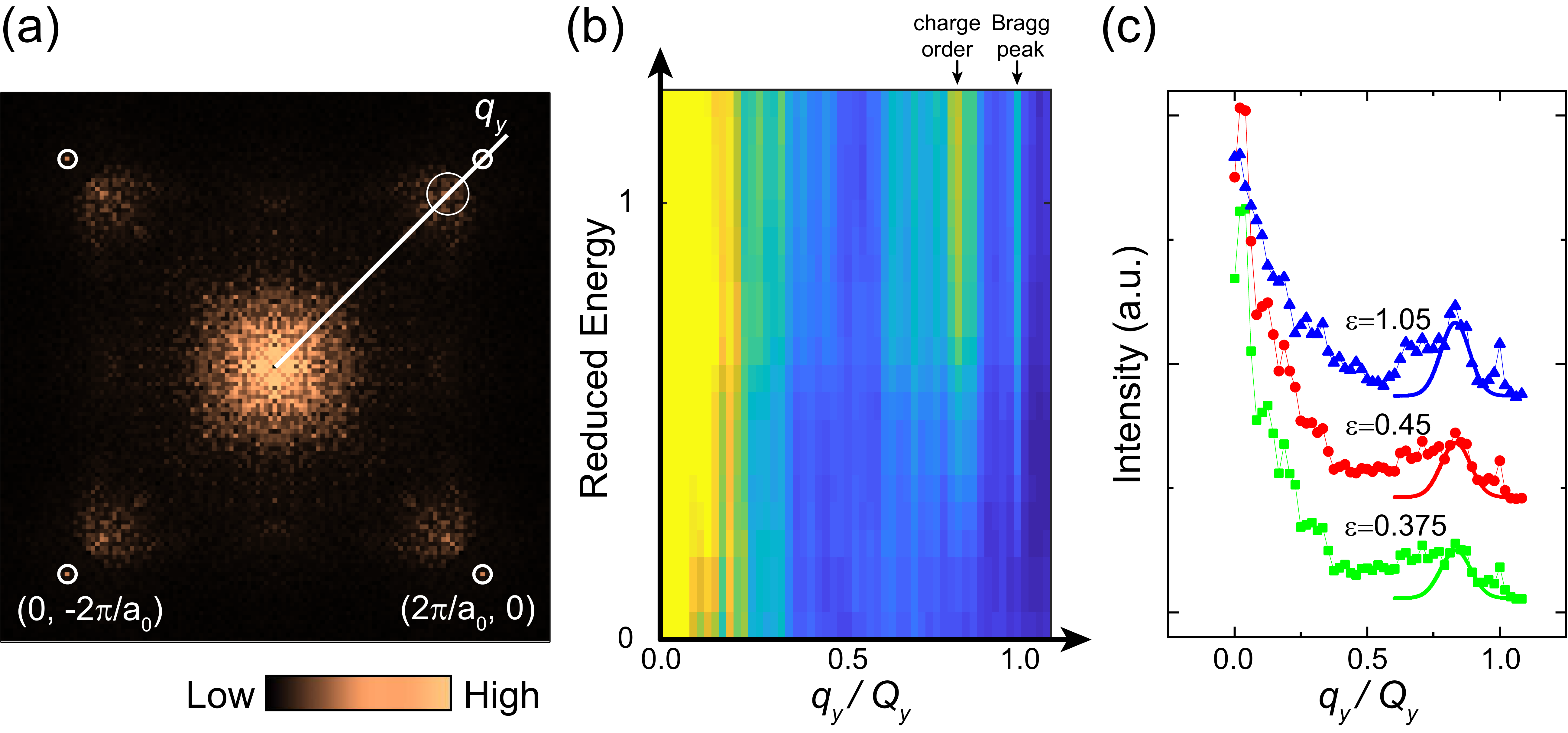}
\caption{(a) The symmetrized ratio map $Z(\boldsymbol{r},\varepsilon=1.05)$.
(b) Along the line shown in (a), linecuts of $\tilde{Z}(\boldsymbol{q},\varepsilon)$ are extracted as a function of
the reduced energy $\varepsilon$.
(c) Three linecuts are shown as examples, with $\varepsilon=1.05$, 0.45 and 0.375. Results of the Gaussian function fitting
are also shown, from which the charge order peak positions are determined.
}
\label{figS2}
\end{figure}

\section{Determining the wavevectors of the charge-order-induced modulation}
\label{secs2}

In Figs.~1(b) and~1(c) of the main text, we present a ratio map $Z(\boldsymbol{r}, \varepsilon=1.05)$ and
its Fourier transform $\tilde{Z}(\boldsymbol{q},\varepsilon=1.05)$ at the PG magnitude. To quantify
the wavevector $\boldsymbol{q}^\ast_{x/y}$ of the charge-order-induced modulation, we symmetrize
$\tilde{Z}(\boldsymbol{q},\varepsilon=1.05)$ with a four-fold rotation symmetry
and a mirror symmetry~\cite{heyangScience}. The signal to noise ratio is enhanced accordingly and the
result is plotted in Fig.~\ref{figS2}(a). A line of $\boldsymbol{q}$ is drawn along the $\boldsymbol{q}_y$-axis
from the center $(0,0)$ to one Bragg peak $\boldsymbol{Q}_y=(0,2\pi/a_0)$ in Fig.~\ref{figS2}(a).
The same operation is applied to the Fourier transform $\tilde{Z}(\boldsymbol{q},\varepsilon)$
at various reduced energies. In Fig.~\ref{figS2}(b), collected linecuts of $\tilde{Z}(\boldsymbol{q}_y,\varepsilon)$ are
shown as a function of the reduced energy $\varepsilon$ and the relative amplitude of the wavevector $|\boldsymbol{q}_y/\boldsymbol{Q}_y|$.
To increase the signal to noise ratio, the pixels of $\tilde{Z}(\boldsymbol{q},\varepsilon)$ close to the $\boldsymbol{q}_y$-axis
are partially added. In addition to the sharp Bragg peak at $\boldsymbol{Q}_y$,
a bright stripe is observed, which trace the peaks induced by charge order.
At each reduced energy, the charge-order-induced peak in $\tilde{Z}(\boldsymbol{q}_y,\varepsilon)$
is fitted with a Gaussian function, as shown by the examples of $\varepsilon=1.05$, 0.45 and 0.375 in Fig.~S2(c).
Centers of the the peak are located at $\boldsymbol{q}^\ast_y\approx0.83\boldsymbol{Q}_y$,
nearly invariant as the reduced energy changes (see Fig.~1(d) of the main text). This result
indicates a static and non-dispersive order.

\begin{figure}[tp]
\centering
\includegraphics[width=0.75\columnwidth]{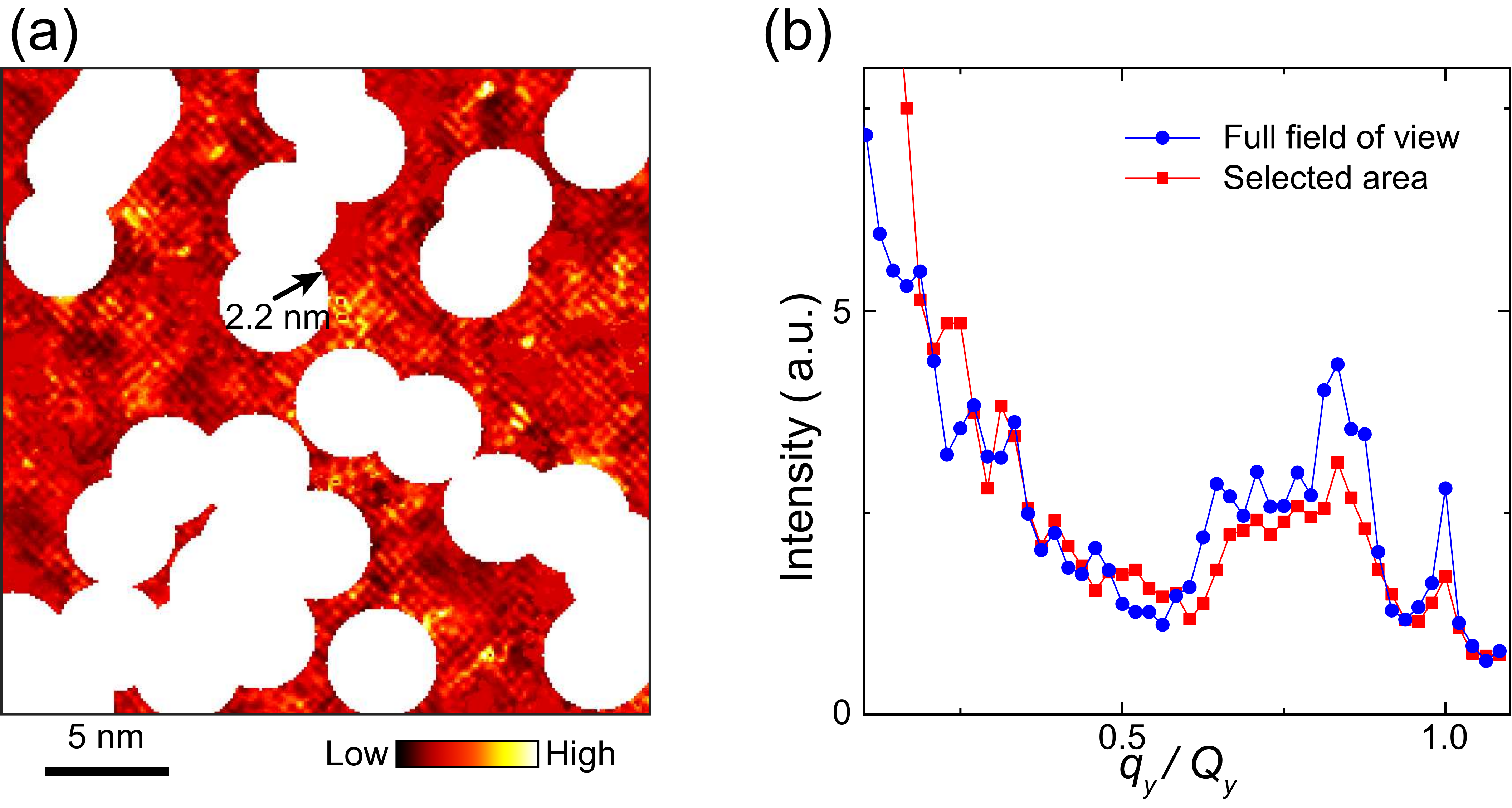}
\caption{(a) The modified map of $Z(\boldsymbol{r}, \varepsilon=1.05)$. A circle is drawn around each defect with a radius of 2.2 nm. A zero value is filled in the selected area.
(b) From the Fourier-transform of the modified real-space map in (a), $\tilde{Z}(\boldsymbol{q},\varepsilon=1.05)$, the linecut
along $\boldsymbol{q}_y$-axis is extracted and compared with that for the full FOV.
}
\label{figS3}
\end{figure}

\section{Influence of topological defects in analysis of the charge-order-induced modulation}
\label{secs3}

Each `topological defect' of the charge-order-induced modulation corresponds to a local disruption of stripe-like modulation
along $x/y$-direction, which also causes a local change to the spatial periodicity. Whether
the average wavevector of modulation in $\tilde{Z}(\boldsymbol{q},\varepsilon)$ is
affected by topological defects is an interesting question to be addressed~\cite{BaggariPNAS2018}.
In the real-space map of $Z(\boldsymbol{r}, \varepsilon=1.05)$,
we draw a circle centered at each defect and with a large enough radius of 2.2 nm. A zero value,
$Z(\boldsymbol{r}, \varepsilon=1.05)=0$, is filled in the selected area as shown in Fig.~\ref{figS3}(a).
Through this modification, the influence of defects is almost completely removed.
Following the treatment in Sec.~\ref{secs1}, we extract the signal of $\tilde{Z}(\boldsymbol{q}_y,\varepsilon=1.05)$ along
$\boldsymbol{q}_y$-axis for the modified map of $Z(\boldsymbol{r}, \varepsilon=1.05)$.
As shown in Fig.~\ref{figS3}(b), the new charge-order-induced peak is located
at nearly the same position as $\boldsymbol{q}^\ast_y$. Thus, the average wavevector
is almost unaffected by topological defects.

\begin{figure}[tp]
\centering
\includegraphics[width=1\columnwidth]{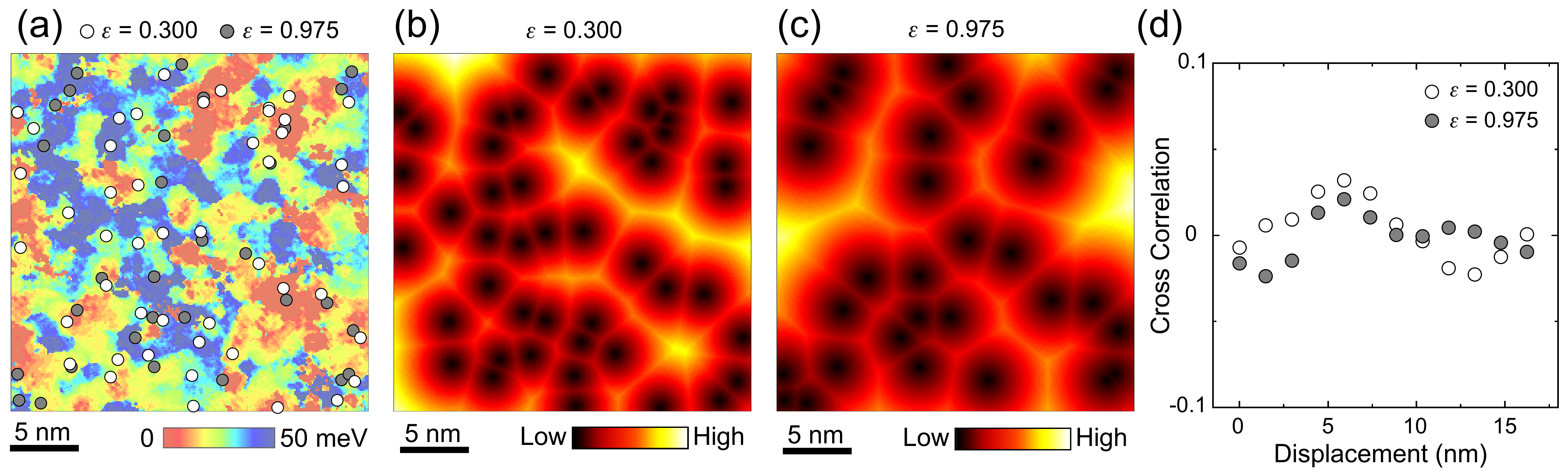}
\caption{(a) The map of the PG magnitude. The topological defects at two typical reduced energies are superimposed by white and black dots, respectively.
(b-c) The distance map of $D(\boldsymbol{r})$ for (c) $\varepsilon=0.300$ and (d) $\varepsilon=0.975$.
(d) The cross correlation functions between the PG and distance maps.
}
\label{figS4}
\end{figure}

\section{Relation between the topological defects and PG}
\label{secs4}
As discussed previously, the PG magnitude $\Delta(\boldsymbol{r})$ can be determined for
each spatial location to produce a two-dimensional PG-map. In Fig.~\ref{figS4}(a), the topological
defects at two typical reduced energies are superimposed on top of the PG-map. No obvious
correlation between the topological defects and PG can be directly observed. We need a further quantitative analysis of the correlation. With the distribution of topological defects, a distance function $D(\boldsymbol{r})$ is defined as the distance between the spatial position $\boldsymbol{r}$ and the nearest defect (Fig.~\ref{figS4}(b) and~\ref{figS4}(c)), similar as in the oxygen defect analysis~\cite{FeiyingSCPMA}. The cross correlation between the PG map and the topological defect distribution is defined by
\be
C(\boldsymbol{R}) = -\frac{\int d^2\boldsymbol{r} [D(\boldsymbol{r})-\bar{D}][\Delta(\boldsymbol{r}+\boldsymbol{R})-\bar{\Delta}]}
{\sqrt{\int d^2\boldsymbol{r}[D(\boldsymbol{r})-\bar{D}]^2  \int d^2\boldsymbol{r}[\Delta(\boldsymbol{r})-\bar{\Delta}]^2  }}
\ee where $\bar{D}$ is the
average distance from the nearest topological defect. An integration over the angular coordinate is further
applied to obtain a cross correlation $C(R)$. As shown in Fig.~\ref{figS4}(d), a negligible cross correlation can be observed in results of $C(R)$ for both the high and low reduced energies.

\begin{figure}[tp]
\centering
\includegraphics[width=0.8\columnwidth]{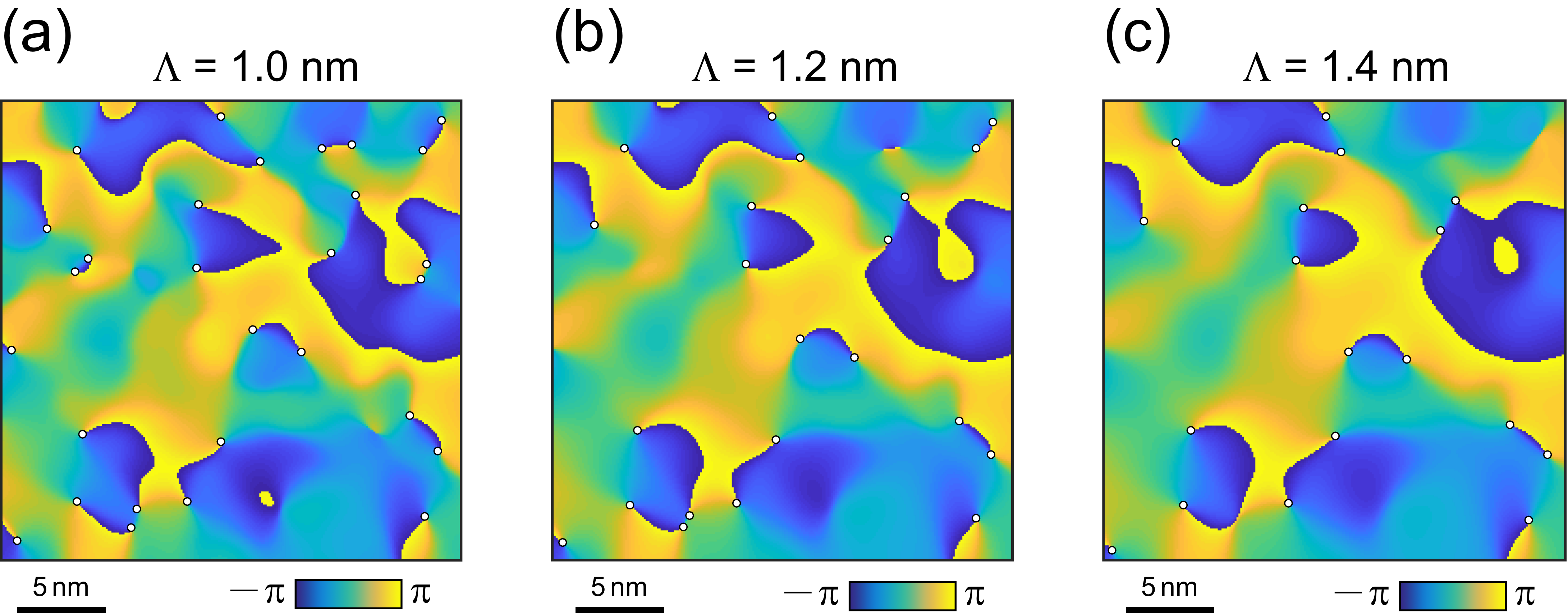}
\caption{Phase $\phi$-maps ($x$-component) extracted from
$Z(\boldsymbol{r},\varepsilon=1.05)$, with the cutoff size $\Lambda=1.0$ nm, 1.2 nm and 1.4 nm, respectively.
}
\label{figS5}
\end{figure}

\section{Effect of the cutoff size in the filtering process}
\label{secs5}

As shown in Eq.~(1) of the main text, a Gaussian function
$\tilde{f}(\boldsymbol{q})=\exp[-(\boldsymbol{q}-\boldsymbol{q}_0)^2/2\Lambda^{-2}]$
is applied to the ratio $\tilde{Z}(\boldsymbol{q}, \varepsilon)$-map, which leads to the $x$- or $y$-component of the modulation
with $\boldsymbol{q}_0=\pm \boldsymbol{q}^\ast_x$ or $\pm \boldsymbol{q}^\ast_y$.
The cutoff size $\Lambda$ is thus a variable parameter in our analysis.
In practice, we choose $\Lambda=1.2$ nm and its inverse $1/\Lambda\approx0.07 \times 2\pi/a_0$
is relatively larger than the standard deviation
of the charge-order-induced peak ($\approx 0.05 \times 2\pi/a_0$) shown in Fig.~\ref{figS2}(c). As a result, the majority
of modulation information is preserved in our filtered maps of $Z^x_f(\boldsymbol{r}, \varepsilon)$
and $Z^y_f(\boldsymbol{r}, \varepsilon)$. To further confirm the reliability of the chosen cutoff size,
we consider two other values, $\Lambda=1.0$ and 1.4 nm, and perform the same filtering process.
The final phase $\phi_x(\boldsymbol{r}, \varepsilon=1.05)$-maps obtained from three cutoff sizes
are plotted in Figs.~\ref{figS5}(a)-\ref{figS5}(c). Despite small changes in the $\phi_x$-patterns
and the number of topological defects, the main feature of the spatial distribution is same
in these three maps. For a relatively smaller cutoff size (or coarsening length), a pair of closely spaced defects is easier to be identified. The smaller the cutoff size, the
larger the number of identified defects. The $\varepsilon$-dependent trend is still kept the same. Thus, the
conclusion from our data analysis is reliable from a qualitative point of view.

\section{Reduced-energy dependence of the spatial distribution of charge-order-induced modulation}
\label{secs6}

\begin{figure}[tp]
\centering
\includegraphics[width=0.8\columnwidth]{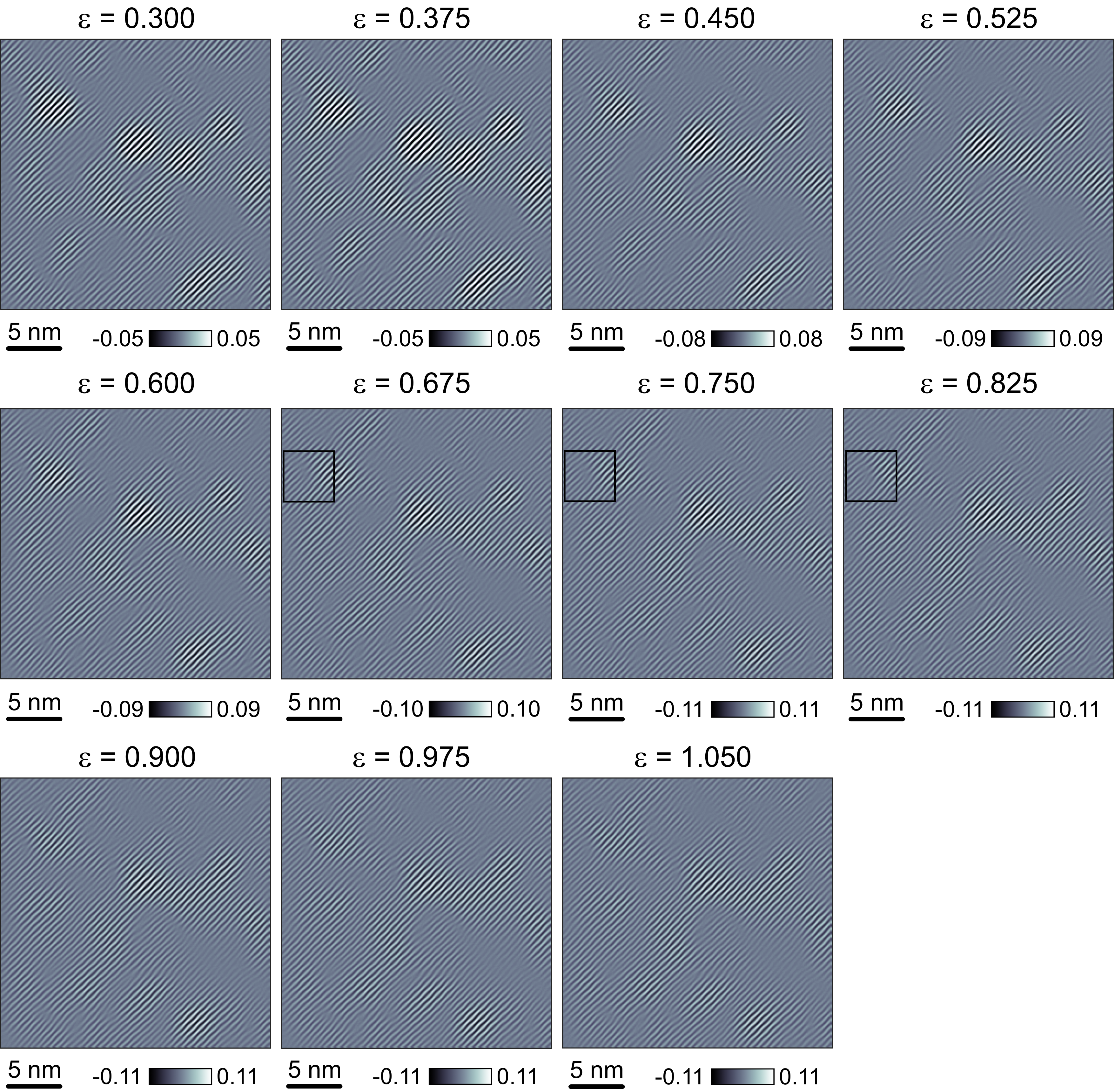}
\caption{ Maps of the charge-order-induced modulation after a Gaussian filtering. For $\varepsilon$ from 0.3 to 1.105, $Z^x_f(\boldsymbol{r},\varepsilon)$
are sequentially shown.
}
\label{figS6}
\end{figure}

\begin{figure}[tp]
\centering
\includegraphics[width=0.8\columnwidth]{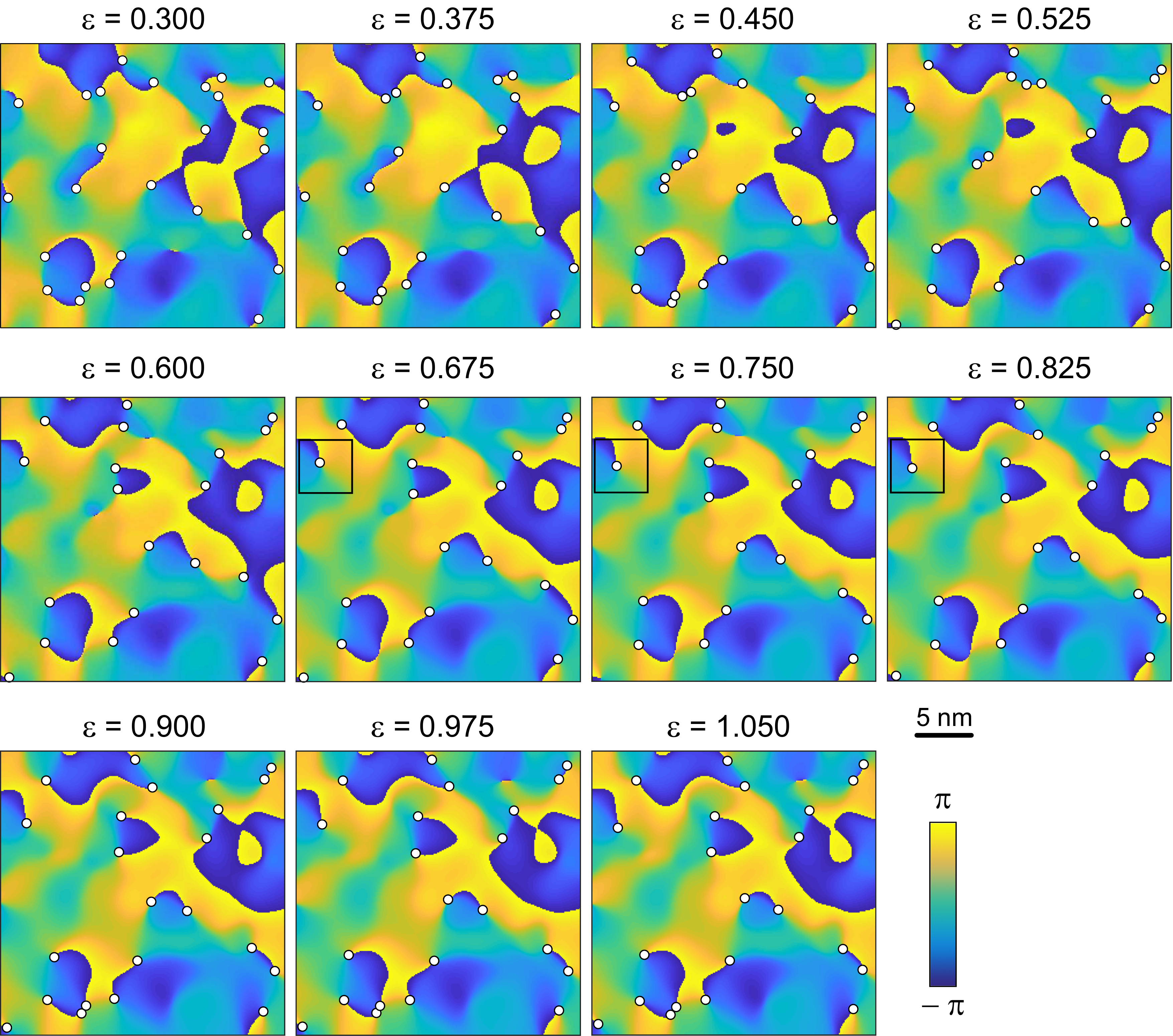}
\caption{Maps of the charge-order-induced modulation after a Gaussian filtering. For $\varepsilon$ from 0.3 to 1.105, $\phi_x(\boldsymbol{r},\varepsilon)$
are sequentially shown.
}
\label{figS7}
\end{figure}

\begin{figure}[tp]
\centering
\includegraphics[width=0.8\columnwidth]{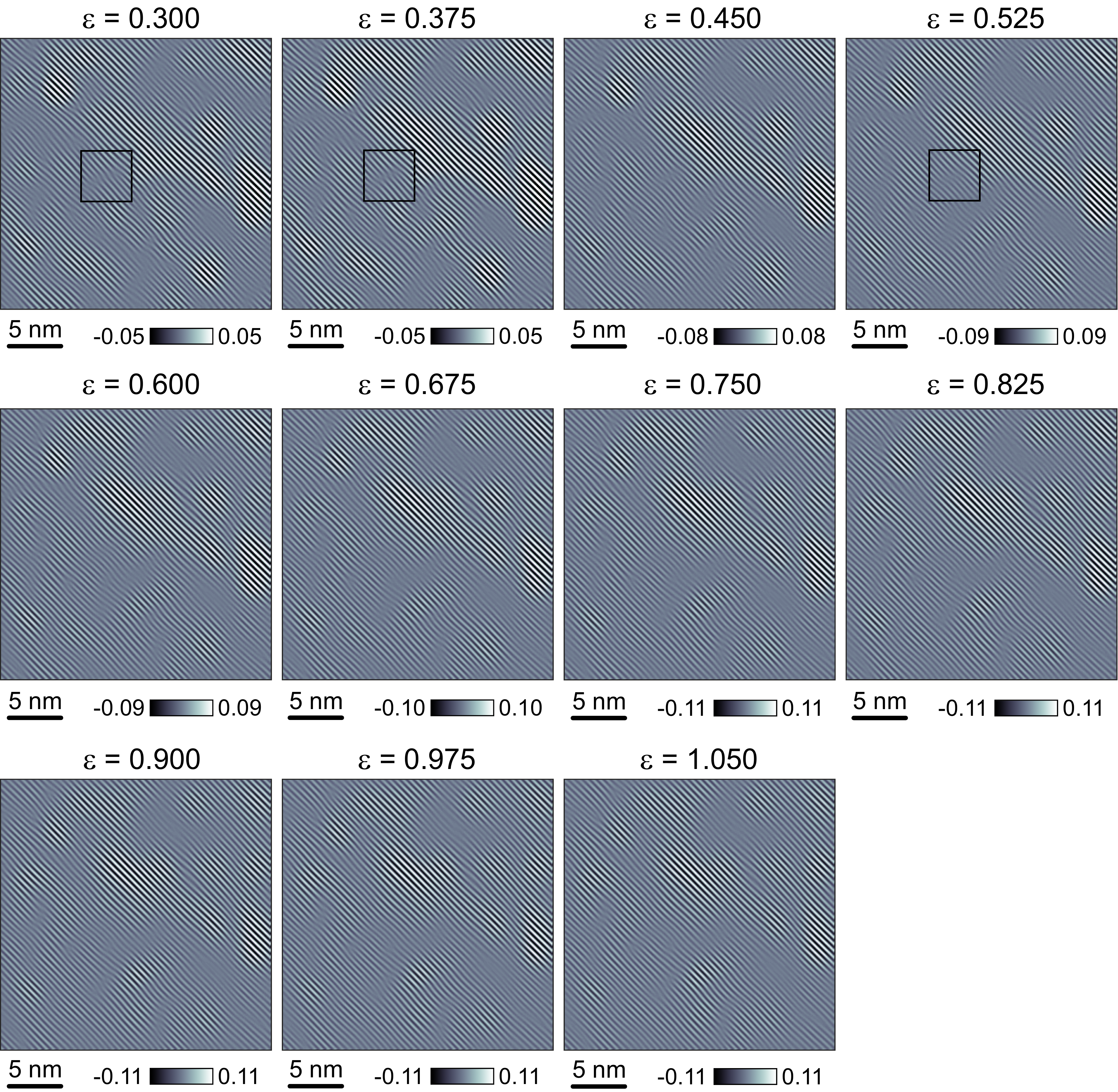}
\caption{ Maps of the charge-order-induced modulation after a Gaussian filtering. For $\varepsilon$ from 0.3 to 1.105, $Z^y_f(\boldsymbol{r},\varepsilon)$
are sequentially shown. }
\label{figS8}
\end{figure}

\begin{figure}[tp]
\centering
\includegraphics[width=0.8\columnwidth]{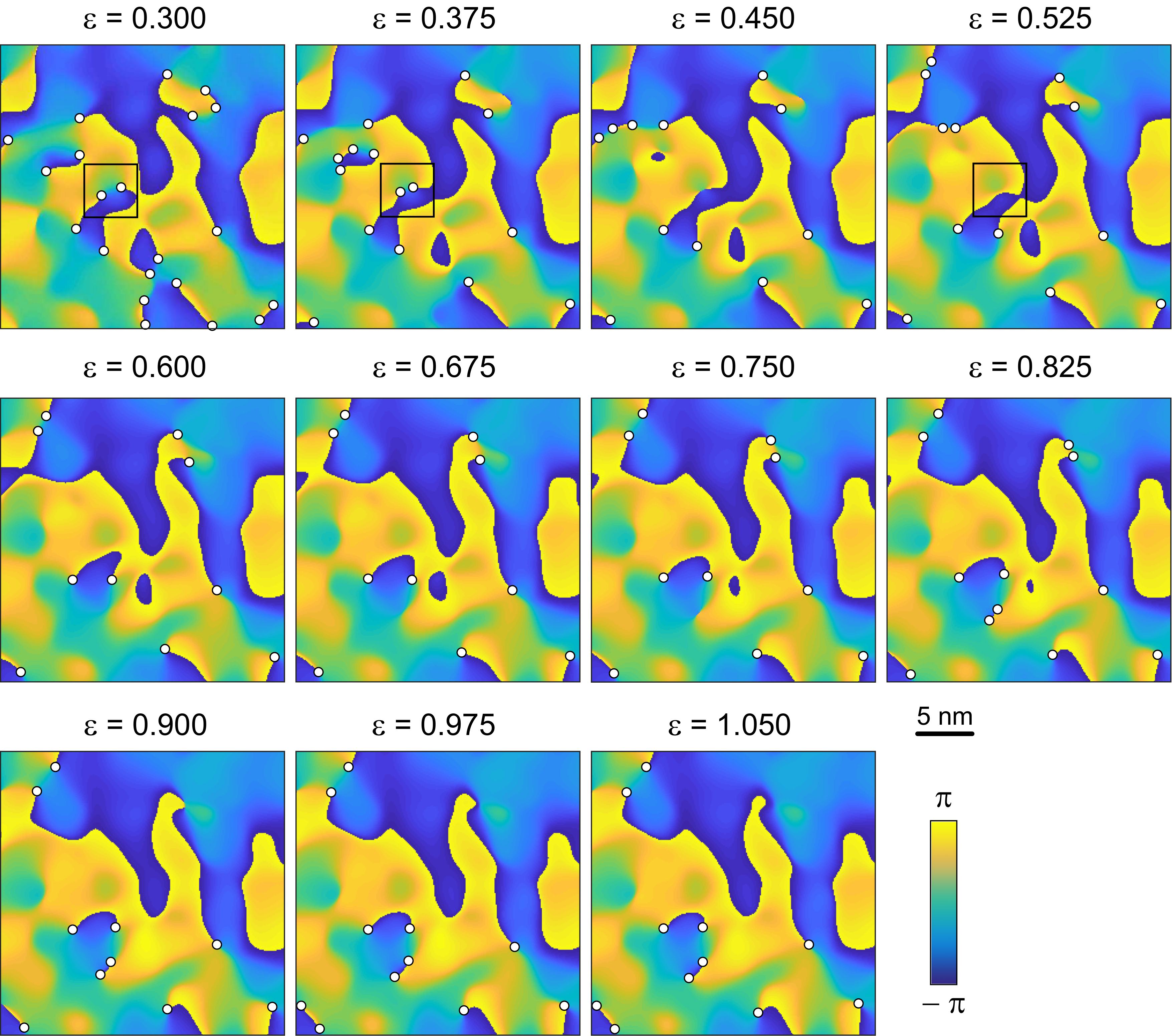}
\caption{Maps of the charge-order-induced modulation after a Gaussian filtering. For $\varepsilon$ from 0.3 to 1.105, $\phi_y(\boldsymbol{r},\varepsilon)$
are sequentially shown. }
\label{figS9}
\end{figure}

In the main text, we discuss the $\varepsilon$-dependence of the spatial distribution of charge-order-induced modulation.
Although the modulation is non-dispersive and the STM is a static measurement tool~\cite{lawler2010intra,fujita2014simultaneous},
its spatial distribution is not necessarily fixed as $\varepsilon$ changes.
In Figs.~\ref{figS6} and~\ref{figS7}, we present $Z^x_f(\boldsymbol{r}, \varepsilon)$
and $\phi_x(\boldsymbol{r}, \varepsilon)$ at 11 different values of $\varepsilon$.
In Figs.~\ref{figS8} and~\ref{figS9}, we present $Z^y_f(\boldsymbol{r}, \varepsilon)$
and $\phi_y(\boldsymbol{r}, \varepsilon)$ at the same values of $\varepsilon$.

\begin{figure}[tp]
\centering
\includegraphics[width=0.8\columnwidth]{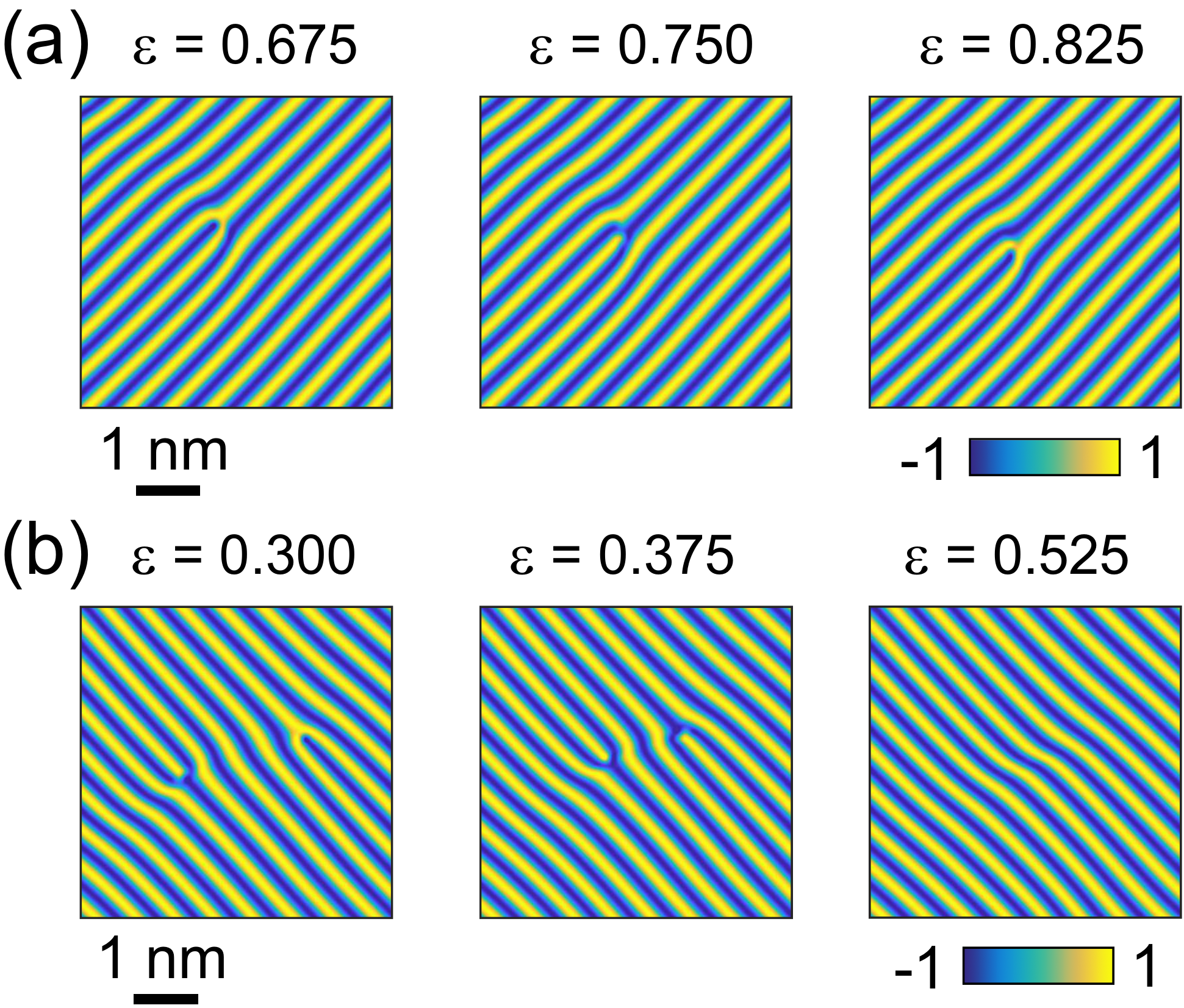}
\caption{(a) An example of how a defect changes position in the stripe structure at different $\varepsilon$. (b) An example of
how two closely spaced defects disappear in the stripe structure at different $\varepsilon$. In both (a) and (b) we present the
normalized maps of $Z^{x/y}_f(\boldsymbol{r}, \varepsilon)/A_{x/y}(\boldsymbol{r},\varepsilon)$ to enhance the signal.}
\label{figS10}
\end{figure}

With the determination of defect locations in $\phi_{x/y}(\boldsymbol{r}, \varepsilon)$, we can map defects
back to $Z^{x/y}_f(\boldsymbol{r}, \varepsilon)$ to visualize the defect (or dislocation core)
`movement' in the stripe structure. Around each defect, the modulation amplitude is close to zero, which obscures
the signal in $Z^{x/y}_f(\boldsymbol{r}, \varepsilon)$. In Fig.~\ref{figS10}, we enhance the signal
by a normalization of $Z^{x/y}_f(\boldsymbol{r}, \varepsilon)/A_{x/y}(\boldsymbol{r}, \varepsilon)$ together with a
cubic interpolation processing. Figure~\ref{figS10}(a) is a typical example of single defect `movement' in
$Z^{x}_f(\boldsymbol{r}, \varepsilon)$ from $\varepsilon=0.675$ to 0.750 and 0.825. Figure~\ref{figS10}(b) is an
example of how two closely spaced defects `annihilate' through stripe bending in $Z^{y}_f(\boldsymbol{r}, \varepsilon)$ from
$\varepsilon=0.3$ to 0.375 and 0.525. The area of these maps are shown by white boxes in Fig.~\ref{figS6}-\ref{figS9}.

\newpage

\newpage

